%% file: thesis.tex
\documentclass[a4paper,12pt,twoside]{report}
\usepackage[left=2cm,right=2cm,top=2cm,bottom=3cm]{geometry}
\usepackage[natbibapa]{apacite}
\usepackage{times}
\usepackage[utf8]{inputenc}
\usepackage{url}
\usepackage[titletoc]{appendix}
\usepackage{float}
\usepackage{rotating}
\usepackage{multirow}
\include{template/thesis.preamble}

\begin{document}

\title{Taxonomies in DUI Design Patterns: A Systematic Approach for Removing Overlaps Among Design Patterns and Creating a Clear Hierarchy
\vskip2cm
\begin{large} 
  Master Thesis
\end{large} 
}

\author{Mubashar Iqbal}
\submitdate{Autumn 2017}

\normallinespacing
\maketitle

\begin{figure}[ht]
  \includegraphics[width=8cm]{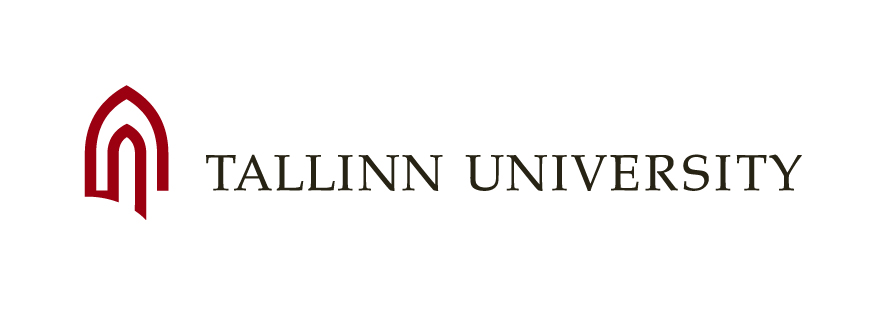}
\end{figure}

\textbf{Non-exclusive license for reproducing the thesis and making it publicly available}
\vskip1.5cm
I, \textit{(name of author) Mubashar Iqbal} \textit{(date of birth) 06.05.1989}
\begin{enumerate}
\item grant Tallinn University a permit (a non-exclusive license) to freely reproduce and make publicly available in the repository of Tallinn University Academic Library a piece of work created by me  

\textit{Taxonomies in DUI Design Patterns: A Systematic Approach for Removing Overlaps Among Design Patterns and Creating a Clear Hierarchy}

supervised by 

\textit{Ilja Šmorgun, David Jose Ribeiro Lamas}

\item I am aware of the fact that the author also retains the rights mentioned in Clause 1. 
\item I certify that granting the non-exclusive license does not infringe on the intellectual property rights of other persons or the rights arising from the Personal Data Protection Act. 
\end{enumerate}

Signed in Tallinn,

\makebox[1.5in]{\hrulefill}\\
\textit{(signature and date)}

\begin{declaration}
I declare that apart from the work whose authors are clearly acknowledged, this thesis is the result of my own and original work.

This work has not and is not being submitted for any other comparable academic degree.
						
The thesis was supervised by Ilja Šmorgun (Tallinn University, Estonia) and David Lamas (Tallinn University, Estonia).

\vskip1.5cm

Mubashar Iqbal

\makebox[1.5in]{\dotfill}(date)

\makebox[1.5in]{\dotfill}(signature) 
\end{declaration}

\begin{dedication}

\centering 
\LARGE This thesis is dedicated to my parents.

\Large For their endless support, love and encouragement.

\end{dedication}


\begin{abstract}
Recently a library of design patterns for designing distributed user interfaces (DUIs) were created to help researchers and designers to create user interfaces and to provide an overview of solutions to common DUIs design problems without requiring a significant amount of time to be spent on reading domain-specific literature and exploring existing DUIs implementations. The current version of the DUI design patterns library need to be assessed because lot of design patterns are overlapping each other and their relationships are not clear enough to effectively find the most relevant design pattern for solving particular design problem, so the purpose of this thesis is to mature the DUI design patterns knowledge field by removing the duplicate design patterns, their description and to create a taxonomy where each design pattern should be organised in a way that will reduce redundancy, possibly leading to grouping or eventually merging similar patterns and allow to navigate to related patterns.

To achieve the defined goals, first target was to investigates the possible overlaps among design patterns and their relevancy with each other, in order to get these insights natural language processing tool was built for extracting and analysing each design pattern research paper to find potential codes. Later in this study thematic analysis was done with domain experts to get themes, their description and higher level categories from generated codes to organize all related design patterns in a clear hierarchy.

The outcomes of this thesis were includes the clarification of the relationships among design patterns by creating a taxonomy, clarified the description of individual design pattern, overlaps and duplicate design patterns were removed and merged similar design patterns.


\end{abstract}
\begin{acknowledgements}
The success and completion of this thesis was required a lot of guidance, support and assistance from many people and all that I have done is only due to such supervision and assistance and I would not forget to thank them.

First of all I would like to express my special thanks of gratitude to my respected supervisors Ilja Šmorgun and David Lamas who gave me this opportunity to work on this project, steered me in the right direction and gave me all the required support and succour that made me to complete this thesis duly.

Secondly, I would like to thank the experts who participated in my research and for their invaluable contribution.

Finally I would like to thank my family, teachers and friends who supported, motivated and encouraged me throughout the study years. Also, I would like to extend my sincere esteems to all the staff members of the School of Digital Technologies and IDLAB.
\end{acknowledgements}

\body

\chapter{Introduction}

The rapid evolution of technology, popularity and diversity of devices has completely changed the way of interaction with interactive systems like multi device environments those support distributed user interfaces to make interaction possible across devices \citep{DelaGuia2014}. 

Such devices can classify as multi purpose devices and specific devices, whereas multi purpose devices refers to the mobile devices and stationary devices. Mobile devices like laptops, tablets, smartphones and smartwatches those are easily available and could be controlled according to the peripheral devices i.e. laptops employ keyboards, mouse and trackpads; and devices like smartphones, tablets, smartwatches employ touchscreens, accelerometers, gyroscopes, GPSs etc; and stationary devices such as Smart TVs, projectors categorized those connects with desktop computers, game consoles and control by novel interaction devices like Microsoft Kinect, the PlayStation Move jointly with the PlayStation Eye or the Wii-mote. Whereas specific purpose devices such as RFID-based panels, plane cockpits, advertisement panels etc are also a part of the available interaction environments to users \citep{Tesoriero2014}.

In the beginning, all these devices were used separately, researchers and designers were unaware of the existence and connection with other devices. In order to carrying out interactive tasks  Distributed User Interfaces (DUIs) have been imagined those completely share the user interface between different interactive devices not just only the information between two different user interfaces \citep{Melchior2011}.

\begin{figure}[ht]
  \includegraphics[width=\linewidth]{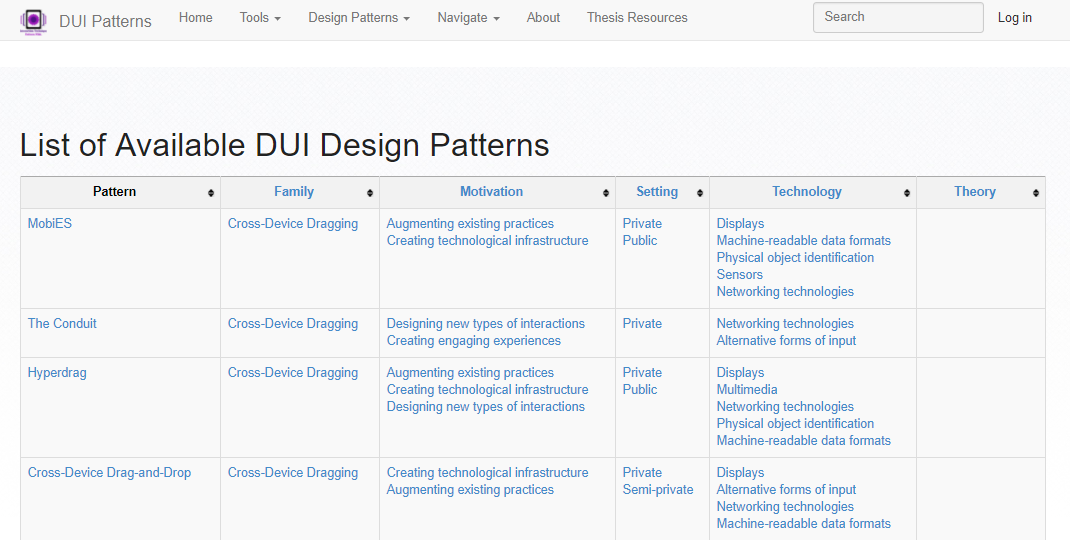}
  \caption{Library of DUI design patterns at semantic MediaWiki}
  \label{fig:DUI_Patterns}
\end{figure}

Recently \cite{Shmorgun2016}, gathered the set of design patterns for aiding the design of distributed user interfaces those are listed at Semantic MediaWiki\footnote{\url{http://idlab.tlu.ee/patterns/index.php/Main_Page}} (see Figure~\ref{fig:DUI_Patterns}). Christopher Alexander originally introduced the term design pattern back in 70’s, as he defined like this: “Each pattern describes a problem which occurs over and over again in our environment, and then describes the core of the solution to that problem, in such a way that you can use this solution a million times over, without ever doing it the same way twice”. Even though this definition was defined for patterns in buildings and towns but what he says is true about object-oriented design patterns \citep{Gamma1994}. And the same way it is applicable on DUI design patterns. According to \cite{Borchers2001}, design patterns provide solution to recurring design problem for aiding to understand problem easily and generate new ideas.

A library of DUI design patterns has been developed to support the process of analysing existing and specifying new DUI design patterns. Currently total 47 design patterns are listed at Semantic MediaWiki, where each pattern is categorised by its name, family, motivation, setting, enabling technology, and supporting theory. The design pattern library as a whole aims to introduce common ways to HCI researchers and practitioners for designing cross-device interactions and to provide an overview of solutions to common DUIs design problems, without requiring a significant amount of time to be spent on reading domain-specific literature and exploring existing DUIs implementations. 

\section{Research Problem}

In current version of DUI design patterns library an issue has been identified that significant number of the patterns are overlapping and their descriptions are not clear enough to be sufficiently useful for design projects. The issue is there because of no clear classification or taxonomy was created to “mature the knowledge field in a way to allows for the description of terms and their relationships in the context of a knowledge area” \citep{Usman2017}. According to \cite{Niu2015}, taxonomy is a knowledge organization system serving as the backbone of the domain knowledge for organizing concepts and applying the findings of a knowledge field \citep{Vessey2005}.

To advance the knowledge and understanding of DUI design patterns there is a need to develop a clear taxonomy where each design pattern should be organised in a way that will reduce redundancy, possibly leading to grouping or eventually merging similar patterns and allow to navigate to related patterns. In another research related to DUI design patterns where \cite{Shmorgun2016} said, there is a need to provide better ways of navigating the patterns collection that would give ease of identification and selection of specific design pattern to researchers and designers for designing DUIs.

\cite{Fincher2000} also focused on taxonomy on theirs four principles of pattern language, where they mentioned; design pattern should have a taxonomy so a reader can find pattern(s) they need, as well as taxonomy descriptions should be clear enough and understandable for people who want to use design patterns for building DUIs in a way that they can distinguish and differentiate among patterns and use them effectively. 

According to \cite{Usman2017}, knowledge classification has supported the maturation of different knowledge fields mainly in four ways:
\begin{enumerate}
\item Classification of a knowledge field eases the sharing of knowledge \citep{SiraVegasNataliaJuristo2009}; and \citep{Wohlin2014}
\item It provides a better way to understand the interrelationships between the objects of a knowledge field \citep{SiraVegasNataliaJuristo2009}
\item Also helpful to identify knowledge field gaps \citep{SiraVegasNataliaJuristo2009}; and \citep{Wohlin2014}
\item Support to make decision process(es) easier \citep{SiraVegasNataliaJuristo2009}
\end{enumerate}

The research problem is that design patterns from DUI pattern library have significant overlaps, their relationships and descriptions are not clear enough to identify and select most relevant design pattern for designing DUIs, so the current version of DUI design patterns library need to be assessed to classify design patterns their knowledge and relationships in a clear structure.

\section{Research Questions}

To achieve the defined research goals first step is to identify the overlaps among design patterns, those overlaps would be removed by performing research strategy and remaining design patterns would be organized by creating a clear hierarchy.

The research questions are as follows:

\begin{itemize}
\item {[RQ1]} Which patterns are overlapping in design patterns library?
\item {[RQ2]} How to remove those overlaps among design patterns?
\item {[RQ3]} How to organize remaining design patterns in a clear hierarchy?
\end{itemize}

\section{Research Goals}

Following research goals were formulated based on the above identified research problem:
\begin{itemize}
\item {[G1]} Remove the overlaps among design patterns
\item {[G2]} Clarify the relationships among design patterns
\item {[G3]} Clarify the description of individual design pattern
\end{itemize}

\section{Foreseen Outcomes}
Following are the expected outcomes based on the above research problem and goals: 
\begin{itemize}
\item Identify the overlaps among different design patterns and provides systematic approach to remove those overlaps. By removing the overlaps it will reduce the redundancy and knowledge of design patterns would be organized based on the similarities those would give clear understanding and ease of identification to researchers.
\item Clarify the relationships among design patterns those will allow researchers to navigate to related design patterns.
\item Clarify the description of individual design pattern that would be useful to advance the understanding and knowledge of design pattern.
\end{itemize}

\chapter{Background}

This chapter discusses important aspects related to distributed user interfaces, design patterns and taxonomy design that serve as a motivation for the research questions those are described in above chapter.

\section{Distributed User Interfaces}

The distribution of user interfaces is a reality \citep{Tesoriero2014}. A distributed user interfaces whose components distribute across several displays of different computing platforms that are used by different users, whether they are working at the same place or remote collaboration. \cite{Elmqvist2011} stated the following five distribution dimensions input, output, platform, space, and time where distributed user interface components are distributed.

\begin{itemize}
\item \textbf{Input (I) -} also called input redirection where either managing input on a single device or distributed across several devices \citep{Myers1998}; \citep{Johanson2002}; \citep{Wallace2008}.
\item \textbf{Output (O) -} also called display or content redirection where graphical output tied to a single display device, or distributed across several devices  \citep{Tan2004}; \citep{Biehl2008}; \citep{Wallace2008}.
\item \textbf{Platform (P) -} the interface executes on a single computing platform, or distributed across different platforms \citep{Elmqvist2011}
\item \textbf{Space (S) -} the interface is restricted to the co-located or remote interactive spaces \citep{Baecker1993}.
\item \textbf{Time (T) -} interface elements execute either simultaneously or distributed \citep{Elmqvist2011}
\end{itemize}

\begin{figure}[H]
  \includegraphics[width=10cm]{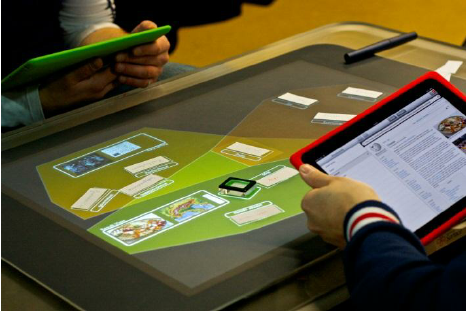}
  \centering
  \caption{TwisterSearch: DUI supporting collaborative web search \citep{Radle2013}}
  \label{fig:Twister_Search}
\end{figure}

\cite{Radle2013} presented TwisterSearch an interactive prototype (see Figure~\ref{fig:Twister_Search}) that was related to a collaborative search system, the main goal of this research was to achieve natural collaboration, collaborative web search and for supporting different working styles on a Microsoft Surface tabletop, Apple iPad tablets, and Anoto digital pen and paper. TwisterSearch was distributed on a shared and private displays where shared display was used to show collaborative search results and private displays were used to publish individual search results and findings on the shared display.

In an another practical work, \cite{Tesoriero2014} created the proxy work system as an example of distributable user interface where he presented the implementation of a set of distribution primitives those are applied to web application environments. Simple HTML based website was created with the navigation bar that was distributed at smartphone and application was distributed on desktop or projected wall, so users were able to navigate through website pages by using the smartphone.

\begin{figure}[H]
  \includegraphics[width=10cm]{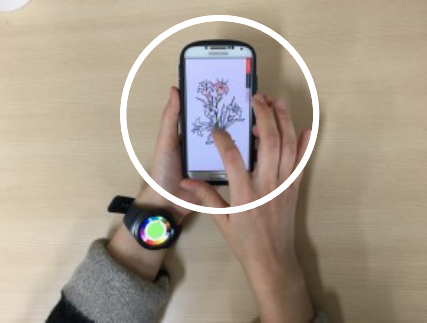}
  \centering
  \caption{TakeOut: Drawing application using DUI \citep{Noh2016}}
  \label{fig:Take_Out}
\end{figure}

Another interesting work was done by \cite{Noh2016} where a drawing application TakeOut (see Figure~\ref{fig:Take_Out}) was created by using distributed user interface. TakeOut drawing application is based on distributed cross device interaction between smartphone and the smartwatch, smartphone drawing application interface distributes on the connected smartwatch where smartphone interface becomes a workspace and smartwatch interface brings tool menus.

\section{Design Patterns}
Design patterns provides reusable solution to a commonly occurring problem, speed up the design and development process, and an effective approach to solve particular problem. This term is mostly familiarly in software engineering, but it exists almost in every field like to address web user interface design, interactive exhibits, user interface related programming, hypermedia applications, or ubiquitous computing, web accessibility \citep{Iacob2011}.

In the same context, \cite{Shmorgun2016} recently created and introduced a library of design patterns in the field of HCI to aid the researchers and designers to support the process of designing DUIs.

DUIs Design patterns are listed at semantic MediaWiki those were collected based on previous studies and from design patterns research papers. Each design pattern is categorised by its name, family, motivation, setting, enabling technology, and supporting theory parameters. Table~\ref{Design_Patterns_Params} is showing the detail of design patterns parameters:

\begin{table}[H]
\begin{tabular}{ |p{5cm}|p{12cm}|  }
 \hline
 \textbf{Parameter} & \textbf{Description}\\
 \hline
 Summary & Each design pattern has summary parameter that extracted from associated research paper that describes about how the design pattern works.\\
  \hline
 Description & A detailed explanation of the pattern that also extracted from associated research paper as it describes what type of interaction technique(s) are involved to make cross device / user interface sharing possible.\\
   \hline
 Design motivation & Describes the primary motivations guiding DUI design pattern.\\
    \hline
 Design goal & In design goal parameter field the main goals are listed those defined to achieve through specific design pattern.\\
    \hline
 Device type & Device type show the context of use either it is private, semi-private or public.\\
    \hline
 Enabling technology & List of technologies that are used for enabling the design pattern.\\
    \hline
 Reference & Reference to the original article where the interaction technique was first described.\\
   \hline
 Pattern family & Each design pattern is grouped with design pattern family.\\
   \hline
 Cites & All articles cited by the one where the pattern is described. \\
   \hline
 Cited by & Where one particular design pattern referred or used as an example to describe an interaction technique.\\
   \hline
 Related to & Show the information of design pattern association with another design patterns either in the context of similar interaction technique, gesture or approach.\\
   \hline
 Examples & A graphical representation of each design pattern to show how it work in real life.\\
 \hline
\end{tabular}
\caption{Parameters of design patterns}
\label{Design_Patterns_Params}
\end{table} 

These parameters are also accounted and studied in this study except summary and description parameters because these parameters were generated from design patterns research papers and these both parameters were already extracted from research papers during the text processing activity.

\section{Taxonomy Design}
Information and knowledge have been classified for centuries \citep{Malafsky2010}. Nowadays, taxonomies are part of our daily life and this is particularly apparent today \citep{Pellini2011}. The concept of taxonomy was originally proposed by Carolus Linnaeus to group and classify organisms by using a fixed number of hierarchical levels. The same concept is adopted now in different knowledge fields, such as education, psychology and computer science etc and different classification structures have been used to construct taxonomies for these knowledge fields \citep{Usman2017}. Taxonomies are a useful and ubiquitous way of organizing information \citep{Chilton2013}.

\begin{figure}[ht]
  \centering
  \includegraphics[width=10cm]{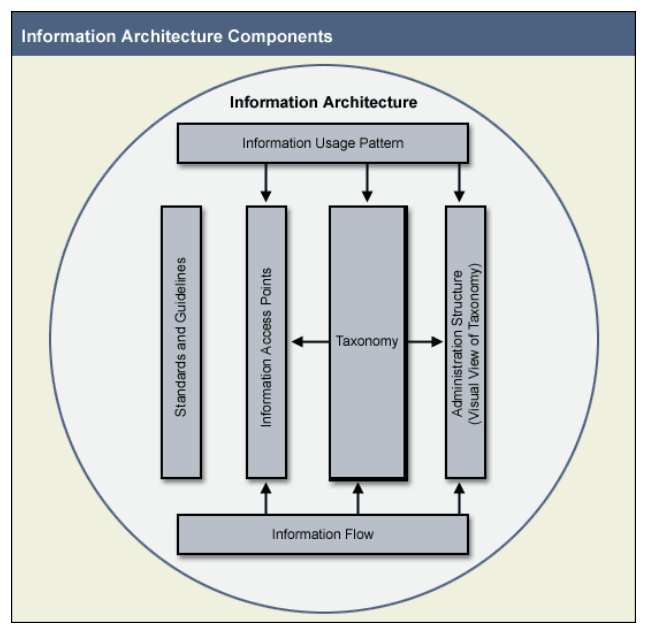}
  \caption{Information architecture components}
  {Source of image: \cite{Tobias2014}}
  \label{fig:Information_Architecture}
\end{figure}

\cite{Tobias2014} illustrated in the diagram (see Figure~\ref{fig:Information_Architecture}) where he showed taxonomy as a core component of information architecture that guides the visual design of information navigation and interrelates with other components.

The advent of the internet in the 1990s contributed an explosion of information dissemination and highlighted the need to develop new tools and skills to organise and retrieve such information. In this context, taxonomies have become necessary \citep{Pellini2011}.

In science and engineering, a systematic description and organization of the investigated subjects helps to advance the knowledge field \citep{Usman2017}. However, little work has been done on taxonomy development in design patterns domain as it is a new but emerging field in the designing of distributed user interfaces. Based on a literature review this study proposes a systematic approach for taxonomy development in the domain of DUI design patterns. 

Based on research and literature review in this study, no systematic mapping or systematic literature review has been conducted to identify and analyse the state-of-the-art of taxonomies in DUI design patterns knowledge field.

This section reviews the definition and purpose of taxonomy, it also includes various set of taxonomies and knowledge classification procedure.

\subsection{Taxonomy Definition and Purpose}

By the definition of Oxford English Dictionary\footnote{\url{http://www.dictionary.com/browse/taxonomy}}, a taxonomy is “a scheme of classification”. Taxonomy word originally came from Greek word taxis, which means ‘order’ and ‘arrangement’. Mostly taxonomy term relates to biology field but in fact everyone use taxonomies in daily life. Every time we enter a modern supermarket, we navigate a carefully studied taxonomy of goods and products located along its aisles \citep{Pellini2011}.

Researchers define taxonomy as a set of structured names and descriptions \citep{Lambe2007a} or controlled vocabulary \citep{Niu2015} that aid to organise data, information and flow of information in a consistent way. 

In daily life knowledge workers spend lot of time for searching and analysing information where large repositories of digital data exists and individuals request to extract exact information what they want. \cite{Serrat2010} said, taxonomy plays an important role to enrich the researchers experience and leverage their expertise, and it only possible when information is well organised and consistent so searching and browsing of information takes less time. \cite{Pincher2010} posits that, all types of management systems are useless without designing a taxonomy for organizing or managing information. 

\subsection{Taxonomy Types}

\cite{Lambe2007a} defined the below structures of taxonomies, specific purposes and when to use.

\begin{figure}[ht]
  \includegraphics[width=\linewidth]{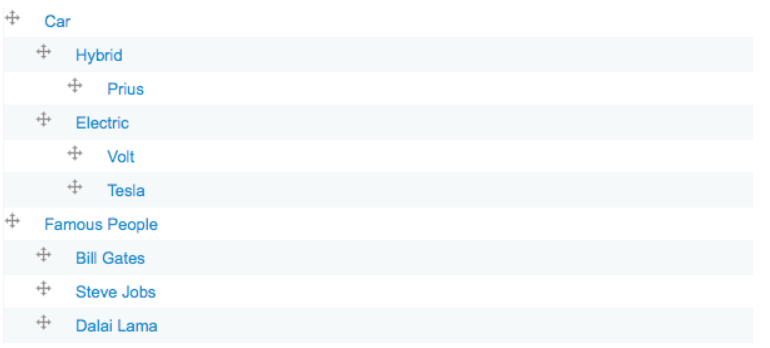}
  \centering
  \caption{List taxonomy}
  \label{fig:List_Taxonomy}
\end{figure}

\textbf{Lists} (see Figure~\ref{fig:List_Taxonomy}) are the more simple form of taxonomy and considered as a good for non-complex issues. Ideally, a list should contain between 12 and 15 elements, but when it becomes longer or more complicated, it is advisable to adopt a different taxonomy form, such as a tree structure \citep{Pellini2011}. 

\begin{figure}[H]
  \includegraphics[width=10cm]{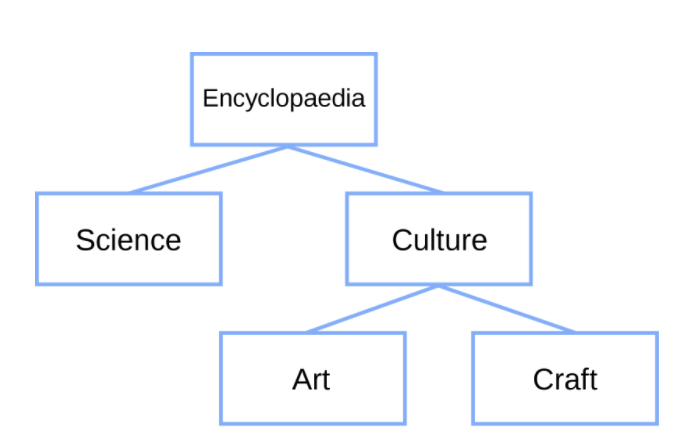}  
  \centering
  \caption{Tree / Hierarchical structure taxonomy}
  {Source of image: Wikimedia}
  \label{fig:Tree_Hierarchical_Structure}
\end{figure}

\textbf{Tree hierarchies} (see Figure~\ref{fig:Tree_Hierarchical_Structure}) are powerful to display cause-effect relationships in the taxonomy and usually use to distinguish broader and generic categories \citep{Pellini2011}. Tree structures are the most used taxonomies and particularly useful when concepts need to be divided into subcategories. They can be represented as pyramidal structure, as this approach mostly used for biological classification.

\begin{figure}[H]
  \includegraphics[width=12cm]{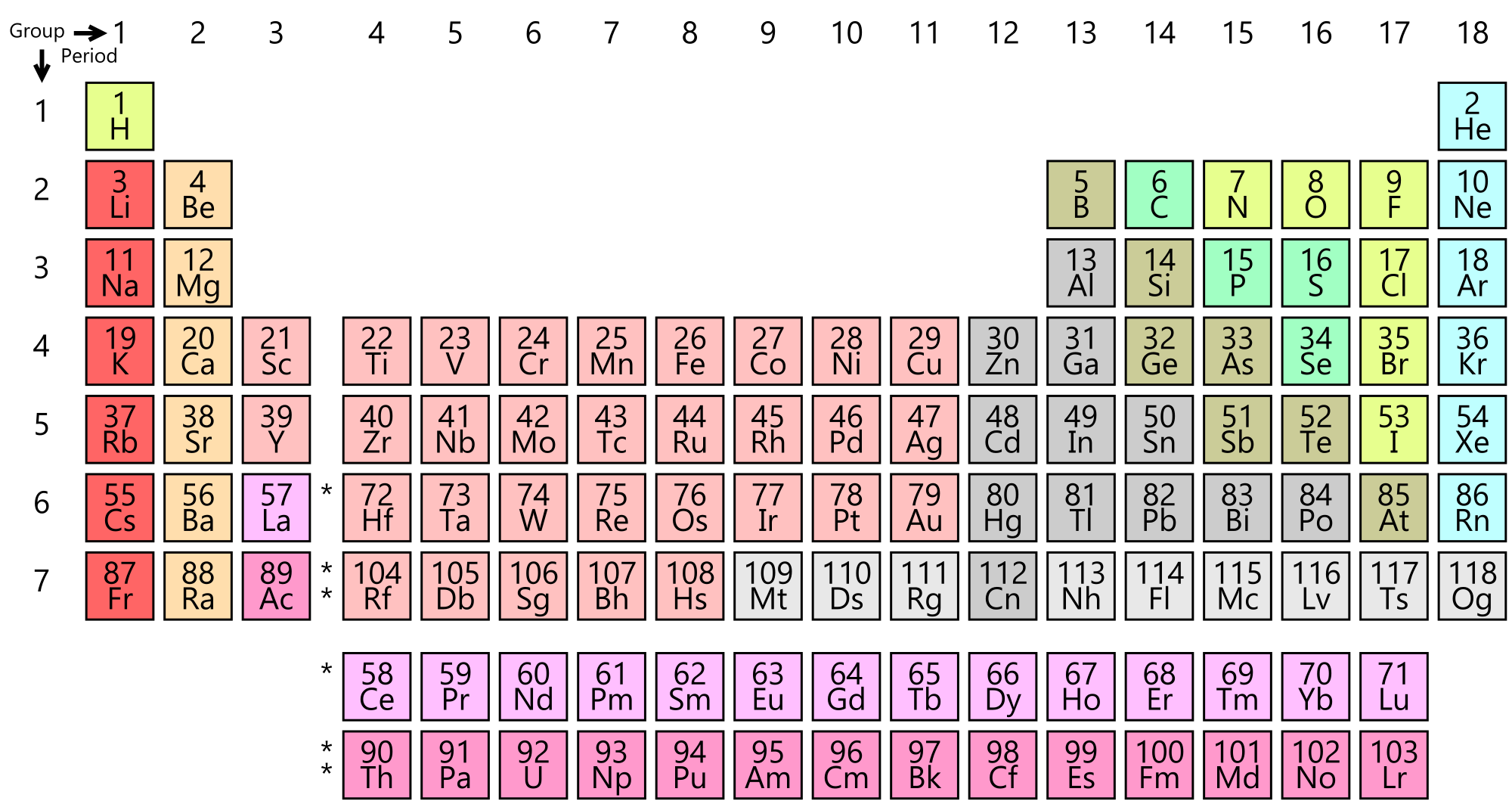}  
  \centering
  \caption{Periodic table: Matrix based taxonomy}
  {Source of image: Wikimedia}
  \label{fig:Periodic_Table}
\end{figure}

\textbf{Matrices} (see Figure~\ref{fig:Periodic_Table}) work best when required to organise information along two or three dimensions. Matrices could be helpful to categories and highlight gaps or missing categories. Mendeleev periodic table of elements is a well known example of two dimensional matrix.

\begin{figure}[H]
  \includegraphics[width=16cm]{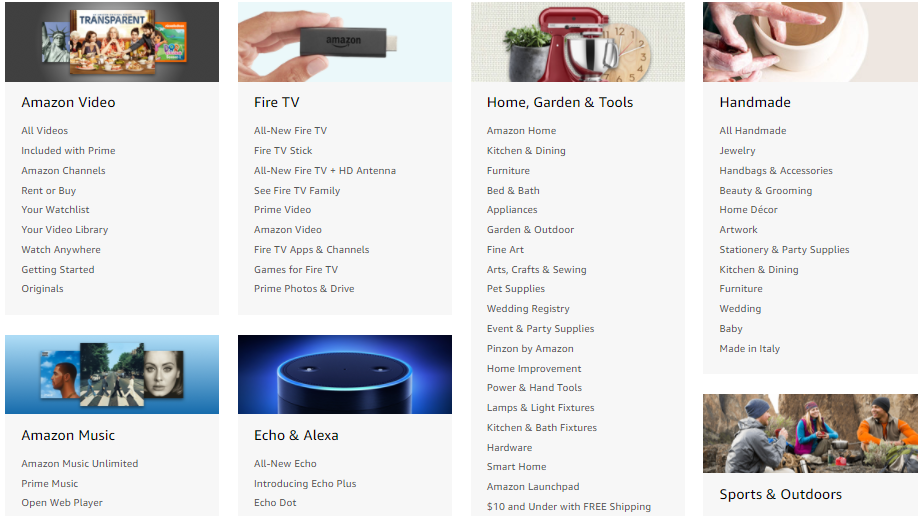}
  \centering
  \caption{Facets taxonomy}
  {Source of image: Amazon.com}
  \label{fig:Facets}
\end{figure}

\textbf{Facets} (see Figure~\ref{fig:Facets}) were introduced first in 1932 by S.R. Ranganathan for classifying books. The basic principle in faceted analysis is that there are more than one perspectives to view and classify a complex entity \citep{Usman2017}. Facets work well with the large content, frequent use of metadata and tags on digital documents \citep{Lambe2007a}. \cite{Pellini2011} said, facets taxonomy is an effective approach when tree structures become too large and complex. 

According to \cite{Pellini2011}, facets mostly use by e-commerce organisations where large publication libraries exists, so the customers can access specific resources and information from different directions easily and effectively. The following screen shot from www.amazon.com where facets taxonomy structure allows the user to find a book by searching through books, audio books, authors, themes, editors, etc.

\begin{figure}[H]
  \includegraphics[width=12cm]{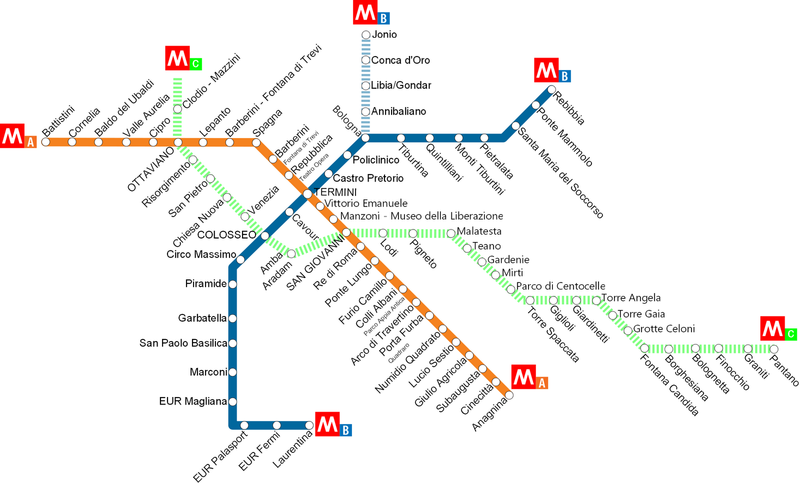}
  \centering
  \caption{System map}
  {Source of image: Wikimedia}
  \label{fig:System_Map}
\end{figure}

\textbf{System Maps} (see Figure~\ref{fig:System_Map}) are visual representations of proximity and connection among categories of a knowledge domain. They are useful when there is a coherent system of knowledge that can be communicated visually \citep{Pellini2011}. System maps are similar to mind maps, and provide a visual representation to show relationships among core ideas \citep{Denham2006}.

\subsection{Subject Matter Expert}
The first step in the design of a new taxonomy is to clearly define the unit(s) of classification \citep{Usman2017}, in this study is a DUI design patterns. 

Lack of subject matter expertise is a disadvantage in taxonomy development. This is understandable and makes sense to assume that the more domain expertise, the better the final information product will be \citep{Earley2008}. Interviewing and conducting analysis with the experts helps to understand where to further explore content in support of taxonomy development. 

This is also an opportunity to prompt them for more specifics about audience needs, asking for example: 
\begin{itemize}
\item Who are your main audiences?
\item Would they understand this term in your opinion? 
\item What are typical user tasks? 
\end{itemize}

\subsection{Methodologies for Building Taxonomy}
Content analysis or textual analysis is often used in finding a taxonomy from a large amount of textual materials \citep{Niu2015}, these methodologies are from social sciences for studying the content of communication, it gained popularity in the 1960s \citep{Niu2015} also \cite{Krippendorff2004} defined content analysis as "a research technique for making replicable and valid inferences from texts to the contexts of their use." By the content analysis approach taxonomy studies mostly focus on the existence of certain words, concept or concepts, later those words and concepts are analysed to find the meaning and relationship with the set of text(s) to speculate the order / hierarchy to classify the information.

\cite{Yao2009} built a tags extraction algorithms for extracting the tags from the tag space of Del.icio.us, where they collected a large set of tags from Del.icio.us to demonstrate the performance of taxonomy construction and evolution. Approximately the collected dataset contains more than 270 million tagging actions from almost 200,000 users in the dated range from 2007.01 to 2008.10. The constructed taxonomy by using this approach reflected the evolution of user interests the organization of online resources and web content.

In an another study, \cite{Chilton2013} presented cascade algorithm technique, a novel crowd algorithm that produces a global understanding of large datasets. Cascade algorithm technique produces human readable category labels and associated items with each category in its decision making algorithm.

\cite{Niu2015} focused on the involvement of domain experts in intensive interviews, iterative development and competency questions use to formulate a taxonomy, these approaches develop taxonomy in a more theoretical sense, those makes the results more convincing and solid as compared to solely using content analysis. \cite{Niu2015} used the ontology annotation tool (OAT) in their study which is widely used for natural language processing (NLP) tasks. After performing data processing, concepts were identified from the text and then each concept was put into the appropriate root concept class. 

\chapter{Text Processing Tool}
\label{Text_Processing_Tool}
This chapter is briefly describing about the software tool that was built to analyse research papers. In order to perform text processing activity web based application was developed that includes different components where each component serving a specific function, those components could be processes, software, hardware, or any other part that helps to built different features of this system.

The technology stack is mostly open source to build this application where I used PHP (recursive abbreviation is Hypertext Preprocessor) an open source web side scripting language that easily embed in HTML to create web pages along with MySQL an open source relational database management system (RDBMS), relational database was created to save required and generated data and information in database for further analysis.MySQL database is easy to install, maintain and integrate with PHP based applications. XAMPP another open source platform that was used as a cross-platform web server solution, consisting apache hypertext transfer protocol server, database and interpreters for PHP programming language. 

Hypertext Markup Language (HTML) was used with Cascading Style Sheets (CSS) and JavaScript to create visual representation for showing data in an organized tabular form to understand and interpret easily. In order to parse and extract text from the PDF based research papers a PHP based 3rd party library name is PDFParser was used. 

\section{Natural Language Processing}
Natural language processing (NLP) mostly used for processing large natural language corpora, with the natural language processing it is an easy and quick way to do text processing and gain insights into what content is being published and how it resonates with the audience, similar like natural language understanding for analysing semantic features like categories, concepts, emotion, entities, keywords, meta data and relations from provided text input. TextRazor, a natural language processing APIs were used to perform text processing activity, it performs deep analysis on content to extract topics, topics relevancy score, entity extraction, concepts extraction and word relations etc.

\section{Selection of Text Processing Service}
A brief systematic review of existing and available language processing services was done before selecting the TextRazor natural language processing services. Spreadsheet was created where entries were listed based on the name of the available service, website URL, short description, capabilities, business model either is it free, trial or paid, supported programming languages, cloud-based or self-hosted or some additional features (see Table~\ref{TextRazor_Systematic_Review} as a reference).

\begin{table}[H]
\begin{tabular}{ |p{6cm}|p{11cm}|  }
 \hline
 \textbf{Name} & TextRazor\\
 \hline
 \textbf{URL} & https://www.textrazor.com/\\
 \hline
 \textbf{Short Description} & The TextRazor API uses natural language processing to extract and understand the concepts and semantics from documents, research, surveys, emails etc.\\
\hline
 \textbf{Capabilities}  & Entity Extraction, Disambiguation and Linking. Keyphrase Extraction. Automatic Topic Tagging and Classification.\\
\hline
 \textbf{Business Model} & 
Charge based on daily requests \textgreater 500\\
\hline
 \textbf{Supported Programming Languages} & REST API, PHP, Java, Python\\
 \hline
 \textbf{Environment} & Cloud hosted, Self-hosted\\
 \hline
\end{tabular}
\end{table}
\begin{table}[H]
\begin{tabular}{ |p{6cm}|p{11cm}|  }
\hline
 \textbf{Additional Features} & Deep analysis,typed dependencies between words and Synonyms. Also offers to increase free limits and special pricing for qualifying academic users.\\
\hline
 \textbf{Limitations} & Paid, no custom model but can create dictionaries and classifications.\\
 \hline
\end{tabular}
\caption{TextRazor natural language processing API}
\label{TextRazor_Systematic_Review}
\end{table}
Following six different services were reviewed based on the above parameters where TextRazor services were selected based on its business model, supported programming languages, environment and importantly concepts extraction capability, whereas other services were either paid, self hosted or did not support concepts extraction. 

\textbf{Google natural language APIs}, derive insights from unstructured text using Google machine learning. It was paid, no custom model and each service charge separately.

\textbf{Microsoft cognitive services}, easily evaluate sentiment and topics from the text to understand what users want. Trial version was available only for 30 days, no custom model and charge by storage and requested services separately.

\textbf{IBM AlchemyLanguage}, is a collection of APIs that offer text processing through natural language processing. It was paid and also required IBM watson studio that cost 150 USD per month..

\textbf{Python NLTK}, is a leading platform for building Python programs to work with human language data. To write own software by using this library that could take several months to develop and also it was self hosting.

\textbf{NlpTools}, same as NLTK, NlpTools is a library for natural language processing written in PHP and to write own software by using this library that could take several months to develop, it was self hosting and NlpTools is still under development and there could be plenty of features missing. \\

Text processing tool was created as a web tool and it is available on the Github\footnote{\url{https://github.com/mubashariq/textanalysistool}} along with their technical documentation at Github wiki\footnote{\url{https://github.com/mubashariq/textanalysistool/wiki/Text-Processing-Tool}}. This system is only accessible by using an internet browser and it also built to performs various actions like exclusion criteria, for helping to perform refinement activity, cross comparisons among design patterns to find commonalities and relevancy with each other.

\section{Limitations}

During building text processing tool some limitations were faced those are mostly related to software side.

\textbf{Secure PDF files} \par
Some PDF files were secured as mentioned above PDFParser does not support secured PDF files to parse and extract data, so manually data was extracted and created .txt files.

\textbf{Invalid PDF structure and character set} \par
Few research papers PDF contains invalid structure and character set so extra work was done in software tool to handle this situation.

\chapter{Research Methodology}
The overall research methodology consisted of three different phases (see Table~\ref{Research_Methodology}). Phase one was mainly focused on the development of text processing tool and data gathering. In second phase of the study was thematic analysis by conducting interview sessions with domain experts. In the last phase of this study was analysis and defining the main category by using potential themes and later organized design patterns in hierarchical structure.

\begin{table}[H]
  \begin{tabular}{ |p{4cm}|p{4.5cm}|p{7.4cm}|  }
  \hline
  \textbf{Stages} & \textbf{Activities} & \textbf{Additional Information}\\
  \hline
  1) Data collection & Generating initial codes & Text processing to find out the initial codes from each research paper by using natural language processing APIs of TextRazor and collected design patterns parameters from semantic MediaWiki. \\
\hline
 \multirow{1}{*}{2) Data Analysis} & Common codes (to answer RQ1) & Initial codes of each design pattern were compared with other design patterns to find the possible duplicates, related terms used for presenting particular design pattern, overlaps among design patterns. \\\cline{2-3}
 & Potential themes (to answer RQ2) & Thematic analysis was done with the domain experts and also social network analysis was performed. Potential themes were identified those helped to formulate root concepts, remove unnecessary overlaps and merging design patterns based on similarities.\\ 
  \hline
  \end{tabular}
\end{table}

\begin{table}[H]
\begin{tabular}{ |p{4cm}|p{4.5cm}|p{7.4cm}|  }
 \hline 
 3) Categorising \& Findings & Organizing in hierarchical structure (to answer RQ3) & Main categories were identified for each design pattern and design patterns were organized in a clear hierarchy.\\ 
 \hline
\end{tabular}
\caption{Research methodology phases and different activities}
\label{Research_Methodology}
\end{table}

\section{Data Collection}

In order to complete the data collection phase, text processing activity was done and design parameters were collected from semantic MediaWiki.

\subsection{Text Processing}
TextRazor were found to be an efficient tool for completing the text processing task by using its natural language processing APIs. In this thesis the main focus was to develop the taxonomies so the first goal was to scout concepts from research papers, in order to achieve the this goal TextRazor API calls were prepared by using parsed text from each design pattern PDF file and sent requests one by one to TextRazor where it processed and response sent back. TextRazor extracted all the information in one request (like entities, topics, topics relevancy score, categories, synonyms, words and words relation and sentences) from where only two parameters "topics" and their "relevancy scores" were selected.

\textbf{Topics} \par
According to TextRazor, it provides an automatic understanding of hundreds of thousands of different topics at different levels of abstraction and useful for tagging to a finite set of categories, or customizing the classification process. The tagging system of TextRazor provides an easy way to add semantic metadata, boost discoverability and textual labels based on provided data text that dramatically reduces the customization effort. 

\textbf{Relevancy Score} \par
The relevance of this topic to the processed document. This score ranges from 0 to 1, with 1 representing the highest relevance of the topic to the processed document. Relevancy score helped to order codes in descending order during the refinement and thematic analysis activities where most relevant codes were listed on top. 

\subsection{Initial Codes}

This phase involves two different stages to gather initial codes, first stage of this phase involves generating initial codes by doing text processing on design patterns research papers text, total 13,834 initial codes were generated and saved in database along with their relevancy scores.

Second stage involves the extraction of design patterns parameters from semantic MediaWiki, parameters were downloaded in JSON (JavaScript Object Notation) format from semantic MediaWiki by using its API commands and later saved in database by using programming, total 1,022 parameters were collected for all 47 design patterns.

\section{Sampling}
DUI design patterns provide a helping source/tool-kit to researchers and designer who want to build interactive user interfaces and displays, so based on the nature of this field and my thesis research goals, the population sample is limited to the domain experts. More specifically domain experts includes from HCI field those participated in different design and development projects, software/quality assurance engineers and designers. \cite{Borchers2001} said, to create successful interactive systems, user interface designers need to cooperate with developers and application domain experts in an interdisciplinary team. Experts profiles were created for clarifying an expert selection criteria to identify eligible participants.   

The purposive sampling (known as selective, or subjective sampling) technique used that involves identifying and selecting individuals or groups of individuals that are especially knowledgeable about or experienced with a phenomenon of interest \citep{Creswell2011}. 

In order to get to rigour, trustworthiness and rich results total 18 participants in the selected sample based on the below created experts profiles (see Table~\ref{Experts_Profile}). To ease the process of thematic analysis and discussions it was an important to know the participant is living in Tallinn city but there was no limitation or restriction to anyone who is willing to participate in this study.

\begin{table}[H]
\begin{tabular}{ |p{3cm}|p{4cm}|p{4.8cm}|p{4cm}|  }
 \hline
 \textbf{Expertise} & \textbf{Status} & \textbf{Education \& Experience} & \textbf{Location} \\
 \hline
  HCI Experts & Currently doing research, studying or working in the field of HCI or interaction design. & Interaction design, human computer or machine interaction, interactive interfaces, distributed interfaces. & Based in Tallinn city or available remotely at Skype.\\
   \hline
  Software \& Quality Assurance Engineers & Currently studying or working in the field of software or quality assurance engineering. & Development of distributed interfaces, interactive displays or building mobile or desktop applications. & Based in Tallinn city or available remotely at Skype.\\
   \hline
   Designers & Currently studying or working in the field of system designing and development. & Designing or development of distributed interfaces, interactive displays or building mobile or desktop applications. & Based in Tallinn city or available remotely at Skype.\\
 \hline
\end{tabular}
\caption{Experts profile}
\label{Experts_Profile}
\end{table}

Total 18 domain experts (see Table~\ref{Participants_Statistics}) participated in thematic analysis sessions where 3 thematic analysis sessions were conducted in group of 2, 2 and 4 experts. Multiple options were provided to approach participants to participate in this study where 3 sessions were conducted at participants home, 3 via Skype, 6 were conducted at participants office and remaining sessions were done in Tallinn University IDLAB.

As per the requirements of thesis, participants age or gender was not important but the educational and professional background was important as it described in above domain experts profiles.

\begin{table}[h]
\begin{tabular}{ |p{10cm}|p{6cm}|  }
\hline
  \textbf{Background} & \textbf{\# of participants} \\
  \hline
 HCI students and professionals & 7\\
  \hline
 Software engineers & 6\\
  \hline
 Quality assurance engineers & 2\\
  \hline
 Designers & 3\\
  \hline
\end{tabular}
\caption{Participants statistics}
\label{Participants_Statistics}
\end{table}

13 out of 18 participants had already knowledge of either distributed user interfaces or design patterns; whereas 10 participants were completely familiar with both terms, and remaining participants were from the similar knowledge field so it was easy to describe and introducing this study to them.

\section{Study Protocol}
Study protocols would help me to follow the plan of action, ensure that the research activities on the track and to protect from the damaging of research activities.

\begin{enumerate}
\item Prior to the arrival of the participant for a session, it is necessary to turn on the computer and set up the design pattern web page (before I created slides but changed after pilot study to show only design pattern MediaWiki Page) and web page would be open in Google chrome browser.
\item Greeting the participant by informal discussions.
\item Hand over the consent form for reading about the sessions and ethics those would be followed during the session.
\item For this step, I created a separate page where I explained about study, goals and its procedure, this page would be hand over to each participant on his/her arrival.
\item Now I would give him/her some time to read about design pattern to understand it and its techniques and will say; ask me if you have any question and let me know when you will finish
\item Participant will read and understand design pattern for 6 minutes (before it was 3 minutes but after pilot study it changed to 6 minutes), but would not tell the exact duration.
\item On this step I will provide web page that contains common codes for one single design pattern, user will read and understand codes.
Then I will ask to generate themes by using listed codes based on his understanding, knowledge and interests. (required time for this step was also changed, before it was 7 minutes and later it changed to 10 minutes)
\item Based on the amount of design patterns (35) I planned to perform total 18 thematic analysis sessions in first iteration of this study where each participant will perform above steps (6-8) two times, so each participant will get two design patterns in each session to find themes from selected codes.
\end{enumerate}

I would inform participant if time goes up to 45 minutes and he can stop any time or finish the task in hand if participant wish to.

The completion of the session is followed by an informal discussion and thanking the user for participating and answering any questions if might they have. And the same procedure will be followed for the second and remaining sessions.

\textbf{Special Case -} In case if I would go to the participant home or working place then step 1 and 2 would not be applicable.

Based on the above activities in study protocol section approximately 40-45 minutes are required to complete one session with two design patterns.

\subsection{Pilot Study}
Research plan and all research instruments for this study were thoroughly checked and pre-tested in a pilot study, that was performed on two participants where one study was completed in 75 minutes and another was completed in 68 minutes. 

Small modifications were done in the time frame for steps 6 and 7 (as mentioned in study protocol section) also participants discouraged to use slides but they preferred to use semantic MediaWiki web page for reading and getting knowledge of design patterns.

Two simple questions were asked only in pilot study to get response from participants based on the employed approach if any further improvements are required.

\begin{enumerate}
\item Is this procedure is time taking, made you tired or bored?
\item Is there any better approach to perform these steps?
\end{enumerate}

One major change was happened based on the responses of above two questions from both participants. Two activities in thematic analysis (reviewing themes, defining and naming themes) were suggested to be completed at home and submit later in defined period of time, as both activities are required some amount of time to think and define. This change was included as an option to ask from the participant either he/she want to finish both sessions completely or wish to complete at home and submit later in defined period.

\subsection{Main Study}
After finishing the pilot study phase with the two domain experts, suggested changes were implemented and session time frame was updated. The study has been carried out in the same way with remaining experts as the pilot study has been conducted. In order to find the right participants, domain experts profile was created as described above and contacted via LinkedIn those were in my connection and also asked from friends to refer your friends those are matched with experts profile.

The time to carry out the study differed for each participant based on individual levels of understanding. Some of the participants had to read the information about the DUIs and patterns multiple times. Also, some of the participants used to ask questions during the study as they encountered misunderstandings. The average duration of the session with each participant was about 45 minutes as planned (excluding two activities, that time in not accounted in this study). 

After completing the session I printed out the codes and created themes and handed over to the participant for performing remaining two activities also time to submit remaining work was defined based on participant wish but was asked to not exceed to more than three days. Also participants were asked to verbally provide any additional feedback about the pattern language and this study. 

\chapter{Data Analysis}
By NLP text processing and design patterns parameters I collected a large corpora of codes, dataset contains 14,856 codes including both initial codes and parameters those were shrunk to 3,406 by applying exclusion and refinement criteria, removed duplicated design patterns codes and accounted relevant codes. Total 18 participants were participated in this study to perform thematic analysis for 35 design patterns.

\section{Common Codes}
This phase involves generating common codes by comparing entire dataset and extracted only relevant codes. After applying below mentioned exclusion criteria on initial codes approximately 13,140 codes were left and 1,716 codes were trimmed off, these trimmed off codes were not removed from the database but status was updated to “deleted”. 

In this stage design patterns were compared with each other design patterns to identify possible duplicates, relevancy and overlaps percentage of each design pattern, total 1081 comparisons were performed for 47 design patterns.

Relevancy percentages were computed based on the similar initial codes, in this activity duplicate design patterns those had 100\% similarity in an initial codes were removed because codes were generated from same design pattern research paper but different techniques were described like HeadLaser, HeadMouse, HandLaser and HandMouse are four different interaction techniques but described in the same research paper. In order to not repeat duplicate codes list for such techniques in thematic analysis only one design pattern technique was used like in this case only HeadLaser was selected.

\subsection{Exclusion Criteria}
\label{Exclusion_Criteria}
In order to clean up the initial list of codes from each design pattern, exclusion criteria was defined to remove the abbreviations, unnecessary and interchangeable terms. Below are the points for exclusion criteria:

\begin{itemize}
\item All those codes were removed who had relevancy score less than 0.078 out of 1.0, by dynamically crawling topics of 47 design patterns research papers where I found least related topic score was 0.079, so this score was defined as a least relevancy limit score. Mostly codes after these were unrelated like abbreviations, words from references etc and has no reason to keep and later use in design patterns topics comparison
\item All those codes were removed from the list by doing manually lookup who had no relevancy either with the topic or design pattern but ranked above 0.079 score (for example topics like Fee, OQO, truth etc).
\item All those codes those were not written in English
\item Mathematical notations, formulas and dates
\end{itemize}

\subsection{Refinement}
\begin{figure}[H]
  \includegraphics[width=\linewidth]{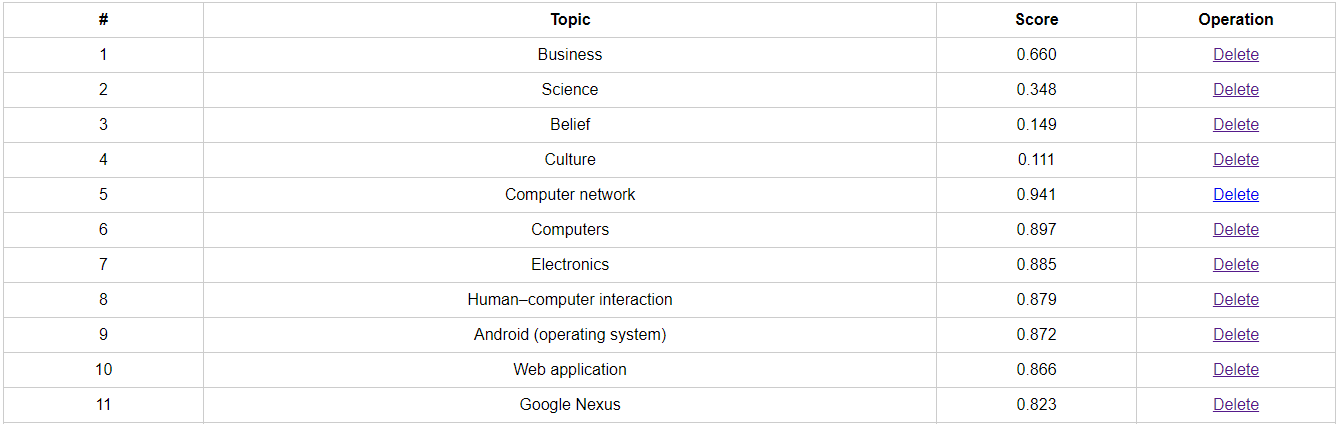}
  \centering
  \caption{Web page of refinement activity}
  \label{fig:Refinement_Activity_Web_Page}
\end{figure}

All general terms those were not suitable for the further development because of already quite abstract and generalize form (i.e. business, belief and culture etc as showed in Figure~\ref{fig:Refinement_Activity_Web_Page}) or interchangeable terms like (interface or interfaces, network or networks etc.). All of such unqualified and generalized terms were trimmed off from the initial codes of each design pattern. 

Refinement activity was done by reviewing one by one each design pattern codes, in order to perform this activity simple web page was created where all identified codes were listed in tabular form (see Figure~\ref{fig:Refinement_Activity_Web_Page}) along with “Delete” button to remove such terms, whereas trimmed off codes were not removed from the database but status was updated to “deleted”. 

After performing the refinement activity and comparison among design patterns total 3,406 common codes were identified from 35 unique design patterns including design patterns parameters. On this stage approximately each design pattern includes 70-130 common codes (see Table~\ref{Design_Patterns_Stats}), those were finalized codes for thematic analysis.

Following Table~\ref{Design_Patterns_Stats} showing an overall statistics of initial codes, common codes and percentage of excluded codes those are related to each individual design pattern.

\begin{itemize}
\item \textbf{Initial codes –} showing the total number of codes those were generated by text processing and gathered design parameters from semantic MediaWiki. 
\item \textbf{Common codes –} common codes where collected by performing two different activities, first was exclusion and refinement activity to clean up the codes and second was cross comparisons among design patterns to find the related codes, it means each design pattern was compared with another design patterns to get common codes.

\item \textbf{Excluded \% –} excluded percentage column in Table~\ref{Design_Patterns_Stats} is giving an information about how much codes were trimmed off by applying exclusion criteria, refinement activity and cross comparisons to consider only related codes.
\end{itemize}

\begin{table}[H]
\begin{tabular}{ |p{6cm}|p{3cm}|p{3.4cm}|p{3cm}|  }
 \hline
 \textbf{Name} & \textbf{Initial Codes} & \textbf{Common Codes} & \textbf{Excluded \%}\\
 \hline
 MobiES & 258 & 89 & 66 \%\\
  \hline
 The Conduit & 460 & 124 & 70 \%\\
  \hline
 Hyperdrag & 318 & 128 & 60 \%\\
  \hline
 Cross-Device Drag-and-Drop & 233 & 89 & 62 \%\\
 \hline
 VisPorter & 375 & 98 & 74 \%\\
  \hline
 MultiSpace & 217 & 99 & 54 \%\\
  \hline
 Conductor & 302 & 96 & 68 \%\\
  \hline
 Cross-Device Pinch-to-Zoom & 312 & 83 & 73 \%\\
  \hline
 Stitching & 327 & 128 & 61 \%\\
  \hline
 EasyGroups & 244 & 88 & 64 \%\\
  \hline
 Bumping & 354 & 110 & 69 \%\\
  \hline
 PaperVideo & 347 & 89 & 75 \%\\
  \hline
 DisplayStacks & 334 & 92 & 73 \%\\
  \hline
 ConnecTable & 327 & 98 & 70 \%\\
  \hline
 Pinch & 261 & 91 & 65 \%\\
  \hline
 Codex & 331 & 102 & 65 \%\\
  \hline
 Interface Currents & 172 & 58 & 61 \%\\
  \hline
 Drag-and-Pop & 259 & 105 & 60 \%\\
  \hline
 Vacuum & 275 & 79 & 71 \%\\
  \hline
 HeadLaser & 288 & 97 & 66 \%\\
  \hline
 Perspective-Aware Interfaces & 337 & 106 & 69 \%\\
  \hline
 Pick-and-Drop & 305 & 129 & 58 \%\\
    \hline
\end{tabular}
\end{table}
\begin{table}[h]
\begin{tabular}{ |p{6cm}|p{3cm}|p{3.4cm}|p{3cm}|  }
  \hline
 Lift-and-Drop & 269 & 98 & 58 \%\\
 \hline
  Slurp & 327 & 105 & 68 \%\\
  \hline
 Ubiquitous Graphics & 309 & 89 & 71 \%\\
  \hline
 SharedViews & 430 & 98 & 77 \%\\
  \hline
 Video Wall & 295 & 87 & 71 \%\\
  \hline
 Chucking & 281 & 98 & 65 \%\\
  \hline
 Voting & 335 & 98 & 71 \%\\
  \hline
 Shuffling & 174 & 84 & 52 \%\\
  \hline
 TranSticks & 338 & 93 & 73 \%\\
  \hline
 Touch-and-Connect & 273 & 83 & 70 \%\\
\hline
 SyncTap & 275 & 94 & 66 \%\\
  \hline
 That One there! & 315 & 92 & 71 \%\\
  \hline
 Select-and-Point & 235 & 99 & 71 \%\\
  \hline
\end{tabular}
\caption{Individual design pattern statistics}
\label{Design_Patterns_Stats}
\end{table} 

\section{Thematic Analysis}
Thematic analysis mostly use in qualitative research, in this study it was perform to examine the codes and collated data to identify significant broader patterns also knows as potential themes. The main reason was behind thematic analysis to find out the users point of view, understandings about the design pattern and how they relate the design pattern information. 

Following activities were performed during the thematic analysis session and in order to follow these defined activities, sample output structure was created (see Appendix~\ref{fig:Thematic_Analysis_Output_Structure}).

\subsection{Searching for Themes}
On this stage I had long list of codes for each design pattern those were identified during common codes analysis, this activity was performed on common codes with domain experts to find the broader level of themes based on the experts point of view, understanding and knowledge. Codes were collapsed to broader level labels and themes, data were visually represented in tabular form to understand the relationship between codes and themes.

\subsection{Reviewing and Naming Themes}
This stage was bind with searching theme stage to refine the themes if possible, whereas some themes were collapsed into other themes and some were break down into smaller components. Later naming themes activity was done in the same stage to get the essence of what each theme is about, effectiveness in the context of relation with particular design pattern and what type of aspects each theme captures. 

\subsection{Defining Category}
Third stage was also performed with the domain experts for defining a main category of each design pattern, it was derived based on the shared information like design pattern example at semantic MediaWiki, codes, created themes and their descriptions. 

After performing the above activities for all total 35 unique design patterns with domain experts data was collected in the format of Appendix~\ref{fig:Thematic_Analysis_Output_Structure}, where created themes section representing the 'Searching Themes' activity,  reviewed, defining and naming themes output was generated by 'Reviewing and Naming Themes' activity and category section is related to the last activity 'Defining Category'.

\section{Data Distribution}

Several steps were performed to arrange the gathered data, first gathered data was read and well understood and then themes were further analysed for cleansing and transforming similar themes, removed unnecessary and irrelevant themes from the list those were not compatible to design pattern knowledge but created by experts based on theirs understanding. In order to perform this process data was extracted from the participants output files and distributed into following three different documents based on the different data.

\textbf{Themes collection} document was created for gathering all design patterns themes and categories from thematic analysis output files. Document structure includes design pattern name, themes and category (see Appendix~\ref{Themes_Collection}), by creating this document it became easy to compare design patterns themes and categories for cleansing and transforming similar themes (i.e. Auxiliary and helper, Mediator and medium etc.), removed unnecessary and irrelevant themes (i.e. Electronic engineering, Hardware etc.); and also all interchangeable themes (i.e. gestural control, gesture control and gesture recognition etc) named to one descriptive term like 'Gesture control', the selection of descriptive term was based on the weighted degree, it means that theme named it which term repeated more.

\textbf{Arranging descriptions}, another document was created for arranging themes and categories descriptions those helped to understand the context and relation of different themes and categories with design patterns. 

\textbf{Interaction variations}, third document was created for listing different design patterns interaction variations (see Table~\ref{Interaction_Variations}), these interaction variations were gathered from design patterns research paper by reading and skimming, these techniques were used by researchers to accomplish one particular design problem in different ways, some techniques showing the different approaches in the design pattern, some shows the variations of different type of devices and displays for making interaction, and some follows the different gesture control.

\section{Themes Social Network Analysis}

\begin{figure}[h]
  \includegraphics[width=15cm]{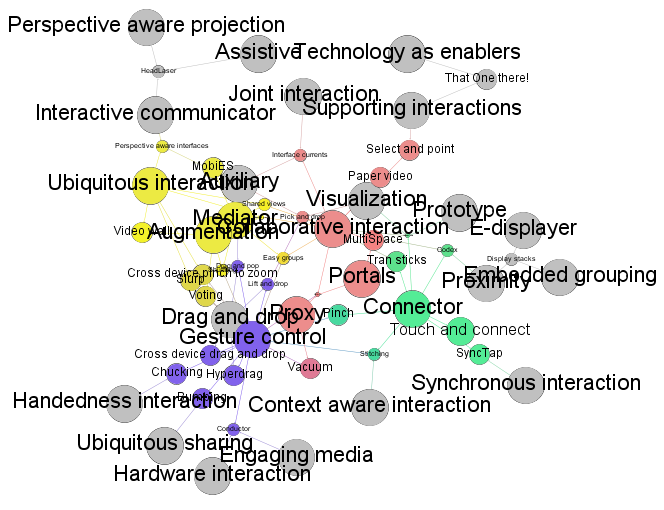}
  \centering
  \caption{Themes social network analysis}
  \label{fig:Social_Network_Analysis}
\end{figure}

Themes social network analysis (see Figure~\ref{fig:Social_Network_Analysis}) was done by using Gephi, an open-source network analysis and visualization software. In order to check how the themes are grouped and to visualize the relationships of different themes towards design pattern. For performing social network analysis data was prepared beforehand by using Google sheet where design patterns theme were listed as a source and design patterns became target. The data was imported in Gephi software and graph was generated by using Yifan Hu force-directed graph algorithm and in-degree partition, Yifan Hu technique considered as a quite easy to understand and effectively shows the weighted nodes of different incoming relationships with edges. Labels were assigned and weighted nodes and their relational edges were coloured those later helped to visualize and understand the relationships of different themes with design pattern that led to generate higher level categories of design patterns.

Graph nodes were scaled according to the in-degree partition and different colours were added to visualize how the themes are grouped together and connections with different design patterns. In graph, 'Ubiquitous interaction' theme and its incoming connections were highlighted in yellow, 'Gesture control' theme and its connections coloured in blue, 'Proxy' themes and its connections in red and 'Connector' theme and its connections represented in light-green colour. These highlighted parts of the graph are clearly giving the clues about the root themes. 

\chapter{Findings}
The previous chapter provided the details of the first two stages of this study where it described how the data was collected and processed. This chapter is covering the third stage for providing the detail information related to findings.

\section{Findings from Common Codes Analysis}
By common codes analysis total 3,406 codes were identified those were later used for thematic analysis. On the same stage design patterns overlaps and duplications were also identified, duplicate design patterns were removed based on the 100\% similarities in initial codes as those were generated from the same research paper. Below Table~\ref{Duplicate_Design_Patterns} is showing the list of 12 different techniques those were excluded based on the 100\% similarities and only one technique was included for thematic analysis. Total 35 unique design patterns were remained after removing such duplicate techniques.

\begin{table}[H]
\begin{tabular}{ |p{6cm}|p{10cm}|  }
\hline
  \textbf{Included Design Pattern} & \textbf{Excluded Design Patterns} \\
  \hline
  Cross-Device Pinch-to-Zoom & Tilt-to-Preview \par
  Face-to-Mirror the Full Screen \par
  Portals \\
   \hline
 \end{tabular}
\end{table}
\begin{table}[H]
\begin{tabular}{ |p{6cm}|p{10cm}|  }
\hline
  HeadLaser & HandMouse \par 
  HandLaser \par
  HeadMouse \\
 \hline
  Drag-and-Pop & Drag-and-Pick \\
 \hline
  Voting & Throwing - MobiComics \par
  Send to me \par
  Retrieving \\
 \hline
  Shuffling & Taking \par
  Throwing \\
\hline
\end{tabular}
\caption{List of duplicate design patterns}
\label{Duplicate_Design_Patterns}
\end{table}

\section{Findings from Thematic Analysis}
By the thematic analysis of 35 design patterns with 18 domain experts, total 89 themes were created along with their description that contains 2-4 sentences and 35 categories one for each design pattern. On this phase various activities were performed to successfully get the desired results from thematic analysis.

\textbf{Searching for themes}, on this stage common codes of each design pattern were analysed and formed either in broader level themes or sub-themes and some codes were discarded those do not belong anywhere. At the end of this stage domain experts created an organized set of themes, approximately 5-10 themes were created for each design pattern.

\textbf{Reviewing and naming themes}, after this activity created themes were concise and immediately give the sense of what the theme is about, also on this step domain experts described each theme in couple of sentences that was helpful to understand the user perspective towards each design pattern that later was used to clarify design pattern description and it also aided to identify the main category of design pattern. After completing this step each design pattern approximately had 2-5 themes along with their description.

\textbf{Defining category}, total 35 categories were created, one for each design pattern in a last activity of this phase that was performed to overall know about the root concepts by relating with design pattern, its codes, created themes and their description. 

\section{Findings from Data Distribution}
From the analysis of data distribution 31 unique themes were identified as showing in following Table~\ref{List_Design_Patterns_Themes}, by going further with these 31 themes analysis 12 design patterns were emerged, selection and naming of design patterns were based on the themes, related research paper, interaction technique and pattern family. As 'drag and drop' pattern was created from 'hyperdrag' and 'cross device drag and drop' themes that was compared with the created themes, the approach followed in these research papers and pattern family for validating the created design pattern name transformation and association with design pattern. In some cases themes were not matching with the design pattern but named after pattern family and implemented approach in research paper like in the case of 'MobiES' and 'The conduit' where the created themes were totally different as compared to pattern family and approach implemented in associated research papers.

\begin{table}[H]
\begin{tabular}{ |p{5.17cm}|p{5.17cm}|p{5.17cm}|  }
\hline
  Acquaintance & Engaging media & Prototype \\
  \hline
  Assistive & Gesture control & Proximity \\
  \hline
  Augmentation & Handedness interaction & Proxy \\
  \hline
  Auxiliary & Hardware interaction & Supporting interactions \\ 
 \hline
 Collaborative interaction & Interactive communicator & Synchronous interaction \\  
 \hline
  Connector & Joint interaction & Technology as enablers \\
 \hline 
 Context aware interaction &  Mediator & Ubiquitous interaction \\
 \hline 
 Drag and drop & Personalization &  Ubiquitous sharing\\
 \hline
E-displayer & Perspective aware projection & Visualization \\
\hline 
Electronic engineering & Pick and drop &  \\
   \hline  
 Embedded grouping & Portals &  \\
   \hline  
 \end{tabular}
\caption{Unique list of design patterns themes}
\label{List_Design_Patterns_Themes}
\end{table}

Generated themes became keywords those would add an extra information in design patterns library to easily navigate among different through them based on their similarities and also it give the essence to understand what design pattern is about.

\subsection{Design Patterns}

Analysis of data distribution brought the following list of design patterns, once the design patterns themes, categories and their relationships have been mapped, it became easy to identify the connected design patterns, 12 unique design patterns were emerged those were grouped based on their similarities in the sense of pattern family and implemented approach in research paper and connected themes.

\begin{enumerate}
\item Drag and drop
\item Drag and pick
\item Lift and drop
\item Bumping
\item Throwing
\item Shuffling
\item Pinch
\item Portals
\item Connector
\item Perspective aware
\item Assistive
\item Proximity
\end{enumerate}

\subsection{Design Patterns Interaction Variations}

The following Table~\ref{Interaction_Variations} is showing a list of interaction variations those occurred in one or more design patterns research papers to solve design problem in a different way or sometime under different names, the list was gathered by reading and skimming the associated research paper of design pattern. These variations were collected to use in taxonomic structures to show the possible variations of particular design pattern or interaction approach.
\begin{table}[H]
\begin{tabular}{ |p{8cm}|p{8cm}|  }
\hline
  \textbf{Design patterns \& Associated Research Paper} & \textbf{Interaction variations} \\
  \hline
 Paper Video & 
 - Spatial Location input \par
 - Gesture displays \par
	\hspace{1cm}- Shake \par
	\hspace{1cm}- Touch \par
	\hspace{1cm}- Quick upward move \par
 - Display proximity input \par
	\hspace{1cm}- Pile \par
	\hspace{1cm}- Side as pointer \par
	\hspace{1cm}- Corner as pointer \\
 \hline
 Interface Currents & - Pools \par
- Streams \\
\hline
 Display Stacks & 
 - Pile \par 
 - Stack and Bend \par 
 - Fan \par 
 - Liner loop \par 
 - Collocation \\
   \hline
Codex & - Concave \par
	\hspace{1cm}- Closed book \par
  	\hspace{1cm}- Book in hand \par
  	\hspace{1cm}- Standing book \par
  	\hspace{1cm}- Lectern \par
  	\hspace{1cm}- Laptop \par
  	\hspace{1cm}- Back to back \par
- Neutral \par
	\hspace{1cm}- Flat \par
- Convex \par
	\hspace{1cm}- Corner to corner \par
	\hspace{1cm}- Face to face \par
	\hspace{1cm}- Battleship \\
  \hline
\end{tabular}
\end{table}
\begin{table}[H]
\begin{tabular}{ |p{8cm}|p{8cm}|  }
\hline
   Lift and drop & - Singleshot \par
- Pick and drop \\
   \hline
\end{tabular}
\caption{Design patterns interaction variations}
\label{Interaction_Variations}
\end{table}

\section{Findings from Social Network Analysis}
Social network analysis was done to check how the themes are grouped and their relationships towards design patterns. Four weighted nodes were identified by running Yifan Hu force-directed graph algorithm along with in-degree partition.


A grouping structure by using Gephi clearly showing root themes and patterns those are gravitated around them. This network analysis approach led to the identification of four new higher level categories by visualizing the central themes and connected design patterns, these central themes became the top part of the taxonomic structure (see Figure~\ref{fig:Proxy_Taxonomy}). 

\textbf{Gesture Control} – Patterns listed under this theme use gestures technique to make cross device connection and moving objects from one interface to another interface, gesture techniques like bumping, pinch, drag and drop etc.

\textbf{Connectable} – all those patterns who use some control like Stylus, TranSticks to connect two different devices.

\textbf{Proxy} – all those patterns are listed here who provide a placeholder on another connected device for sharing information.

\textbf{Ubiquitous Interaction} – design patterns those enable cross device communication without any physical barrier (also called transparent interaction technique) and by adaptive interfaces.

\section{Taxonomic Structure}

Based on the above research findings four hierarchical taxonomy structures were generated where design patterns inherits with a higher level pattern category. Taxonomy structures are divided into four different levels where it structured by category, families, design patterns and interaction variations. Following are the taxonomic structures levels, where:
\begin{itemize}
\item \textbf{Level 1}, is shaded in blue colour where main category is listed. \item \textbf{Level 2}, grey colour boxes are design patterns families and representing level 2 those were generated in one previous study of \cite{Mercer2015}, total 9 families are listed at semantic MediaWiki and patterns were initially grouped in these families. 
\item \textbf{Level 3}, is highlighted in pink colour and it represents design patterns those are newly created in this study. 
\item \textbf{Level 4}, red colour boxes representing different design patterns interaction variations those were implemented in different research papers and collected in this study.
\end{itemize}

\begin{figure}[ht]
  \includegraphics[width=\linewidth]{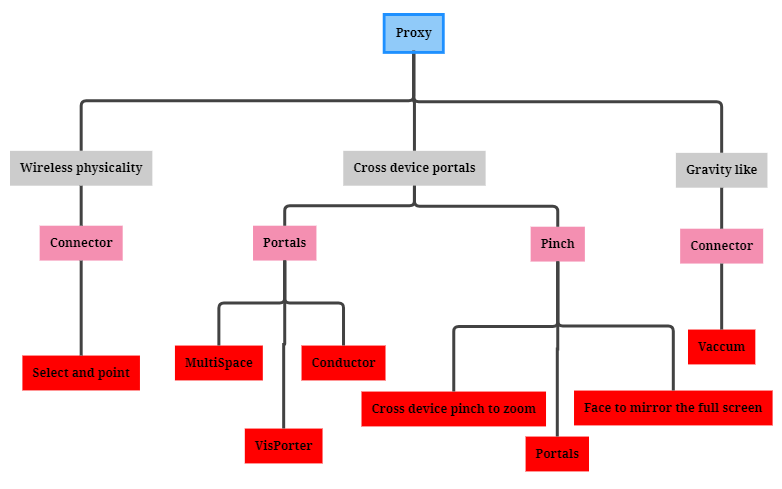}
  \centering
  \caption{Proxy taxonomy structure}
  \label{fig:Proxy_Taxonomy}
\end{figure}

Listed below taxonomic structure (see Figure~\ref{fig:Proxy_Taxonomy}) contains 3 different families and 4 design patterns those are categorised based on the category and families. Like in the example of 'Cross device portals' family where 'Portals' design pattern is listed based on the their shared similarity and interaction technique. Each design pattern has different implementations or interaction variations those are listed in below design pattern in red colour boxes.

\chapter{Discussion}
The aim of this study was to remove the overlaps among design patterns and clarifying their relationships that led to structure design patterns library in a clear hierarchy. The taxonomic structure was necessary as it is a backbone of any knowledge filed \citep{Niu2015}, as well as it helped to understand the relationships among design patterns, so the researchers can easily navigate through design patterns library.

Systematic approach was followed to advance the knowledge and understanding of DUI design patterns by creating a clear taxonomy where each design pattern is organised in a way that reduces the redundancy, led to grouping and merged similar patterns. In one study \cite{Shmorgun2016} said, there is a need to provide better ways of navigating the patterns collection that would give ease of identification and selection of specific design pattern to researchers and designers for designing DUIs, so now this study clarified the relationships of design patterns those allowed to navigate to related patterns easily.

Following research questions were set to answer the identified research problem and achieving the desired goals.

\begin{itemize}
\item {[RQ1]} Which patterns are overlapping in design patterns library?
\item {[RQ2]} How to remove those overlaps among design patterns?
\item {[RQ3]} How to organize remaining design patterns in a clear hierarchy?
\end{itemize}

In order to answer these research questions, various research activities were performed (see Table~\ref{Research_Methodology}), where first activity was performed for data collection, second activity was done by doing thematic analysis \& finding potential themes and last activity was related to categorization and findings.

To answer the {[RQ1]}, initial codes were generated by using text processing tool (see Chapter~\ref{Text_Processing_Tool}), the tool was built by myself in this study to extract core concepts from design patterns research papers, initial codes were compared to identify possible duplicates and overlaps of design patterns with each other, total 12 duplicate patterns were identified based on the 100\% similarities in initial codes.

The aim of the second research question was to remove the overlaps so to answer the {[RQ2]}, thematic analysis sessions was conducted with domain experts (see Table~\ref{Experts_Profile}), total 18 experts were participated in this study to perform thematic analysis for 35 design patterns. 31 unique potential themes were identified for each design pattern based on the experts output those helped to formulate root concepts, remove unnecessary overlaps and merging design patterns based on their similarities. 

To answer the last research question {[RQ3]}, main categories were identified for each design pattern and design patterns were organized in a taxonomic structure. Only four new higher level categories were created and 12 design patterns were identified, 9 design patterns families were collected from one previous study of \cite{Mercer2015}), on the same stage different interaction variations (see Table~\ref{Interaction_Variations}) were collected as well from design patterns research papers.

Research goals and expected outcomes were achieved effectively, duplicate were removed, systematic approach was followed to remove overlaps from design patterns, reduced the redundancy and knowledge of design patterns is now organized in a taxonomic structure. Also description and relationships of design patterns now clarified, no duplication exist and also added new knowledge in the design patterns library by adding themes as a keywords those give an essence to understand what design pattern is about.

Current study has few limitations as well those could be improved in future. One possible limitation was that only two participants were from designing distributed user interfaces field, also there was no working prototype or implemented example of design patterns to educate participants. Due to these reasons it was difficult to explain and educate participants particularly in the DUIs design patterns field. 

Second limitation is that newly created taxonomies are not validated and without this I could not claim these provide better understanding, ease of identification and selection of design patterns. Another limitation was that taxonomies were developed without taxonomy development expert(s), so by overcoming this shortcoming would reveal more stronger and interesting results. 

\chapter{Conclusion}

The aim of this research was to classify DUI design patterns library and clarify the relationships of design patterns by removing the duplicates and overlaps among them. The proposed systematic taxonomy development approach was doable to guide the taxonomy development process, and constructed taxonomy reflected the evolution of experts understanding and knowledge, design patterns content, and facilitate the organization of design patterns in taxonomic structure.

This paper presents a text processing tool that was built by using an open source technology stack. Text processing tool used the TextRazor natural language processing services that efficiently generated the core concepts / initial codes from the design patterns research papers. Common codes were identified by comparing design patterns with each other, on this stage duplicate design patterns techniques were identified and excluded them before moving to further analysis. Unique design patterns common codes were further analysed by thematic analysis where 31 themes were identified those were later added in design patterns library as a additional knowledge to easily navigate among different design patterns based on their similarities and give the essence to understand what design pattern is about.

It was also important to see how the themes are grouped and connected with each other and design patterns, social network analysis was done by using Gephi software and graph was generated by using Yifan Hu force-directed algorithm and in-degree partition. By performing this data analysis activity four new higher level categories were created. Data distribution analysis was also done to understand the meaning of the collected data and formulate the information from thematic analysis data, unique design patterns list were identified and later design patterns were organized in clear taxonomic structures. 

The results produced this approach were more competitive in quality and time effective, as it uses NLP text processing services to extract core concepts from the provided text so work efforts reduces significantly. 

Furthermore, although as noted previously, the methodology has some limitations and areas to be explored in future, it can be concluded that this study successfully presented the classified taxonomic structures of design patterns library by following the systematic taxonomy development approach that could also be helpful in future to classify new design patterns or related knowledge in a particular taxonomy.

\chapter{Future Work}
Outcomes of this study has opened several new research areas to advance the DUI design patterns knowledge field. One possibility to conduct a research with end users to validate newly created taxonomies, as \cite{Usman2017} said; validation strengthens reliability and usefulness of taxonomies; and it would also be informative to investigate about how users navigate to the design patterns library after implementing new taxonomic structure without encountering excessive cognitive load to find the desired design pattern for specific design problem solution.

Another possibility for the future work could be describing and detailing the design patterns keywords those added newly in design patterns library to give more thorough overview of the patterns, identify and map alternate examples and their connections so the researchers would easily understand the patterns better and use them more effectively. As \cite{Pellini2011} said, keywords represents as folksonomies, folksonomies are not an alternative to taxonomies but help to enhance agility, awareness, shared understanding and linked meanings.

\chapter{Kokkuvõte}
\textbf{Hajutatud kasutajaliideste disainimustrite taksonoomia: süstemaatiline lähenemine kattuvuste eemaldamiseks ja selge hierarhia loomiseks.}

Hiljuti loodud hajutatud kasutajaliideste disainimustrite kogu aitab teadlastel ja disaineritel luua kasutajaliideseid ning annab ülevaate lahendustest tavapärastele hajutatud kasutajaliideste disainiprobleemidele, ilma et oleks vaja kulutada palju aega selleks, et lugeda valdkonna kirjandust ja uurida olemasolevaid hajutatud kasutajaliideste teostusi. Olemasolevas hajutatud kasutajaliideste disainimustrite kogu versioonis on kerkinud probleem, et märkimisväärne hulk disainimustreid kattub ja nende kirjeldused ei ole piisavalt selged, et olla küllaltki nii kasulikud disainiprojekti kõige olulisemate disainimustrite tuvastamiseks ja valimiseks. Probleem seisneb selles, et puudub selge klassifikatsioon, mis võimaldaks disainimustrite konkreetsesse järjekorda seadmist.

Käesoleva magistritöö eesmärgiks on edendada hajutatud kasutajaliideste disainimustrite teadmiste valdkond, kõrvaldades korduvaid ja kattuvaid disainimustreid. Selgitades nende kirjeldust ja luues taksonoomia, kus iga disainimuster on liiasusi vähendavalt korraldatud, tuua kaasa samalaadsete mustrite rühmitamise ja ühendamise, et liikuda seotud mudeliteni.

Lõputöös esitatud eesmärkideni jõudmiseks jaotus töö etappideks, millest esimene oli uurida võimalike kattumisi disainimustrites ja seotust nende vahel. Uuringu raames oli võetud aluseks tekstianalüüsi vahend, võtmaks välja ja analüüsimaks iga disainimustri kirjeldust ja leidmaks võimalike märksõnu, mis kõige paremini seda disainimustrit kirjeldavad. Keelepõhise tekstianalüüsi vahendi loomiseks kasutati TextRazor nimelist teenust ja vahend ehitati, kasutades avatud lähtekoodiga tarkvara tehnoloogiat.

Teine etapp eeldas tekkinud märksõnade puhastamise ja selleks, et asjatud ja tähtsusetud märksõnad eemaldada, määratleti välistamiskriteeriume. Samuti oli teostatud viimistlus abstraktsete ja üldiste märksõnade kõrvaldamiseks. Viidi läbi temaatiline analüüs domeeniekspertidega, et saada kõrgema tasemelisi ja informatiivseid teemasid. Nende kirjeldustest ja kategooriatest loodud märksõnu ning teemasid omakorda paigutati ning analüüsiti, kujundamaks lõpuks kõrgetasemeline kategooria, mis on vajalik kõikide seonduvate disainimustrite korraldamiseks selge hierarhiana.

Käesolev lõputöö täitis kõik eelnevalt püstitatud eesmärgid ja hõlmab iga disainimustri hierarhilist struktuuri. Selgitades nende omavahelisi suhteid, sisaldab see iga konkreetse disainimustri jaoks seonduvaid märksõnu, mida loodi inimese ja arvuti interaktsiooni ekspertide arusaamal. Kattuvad ja korduvad disainimustrid eemaldati, sarnased disainimustrid ühendati ning liigitati ühe termini alla.

\appendix
\chapter{Thematic Analysis Output Structure}
\begin{figure}[ht]
  \includegraphics[width=12cm]{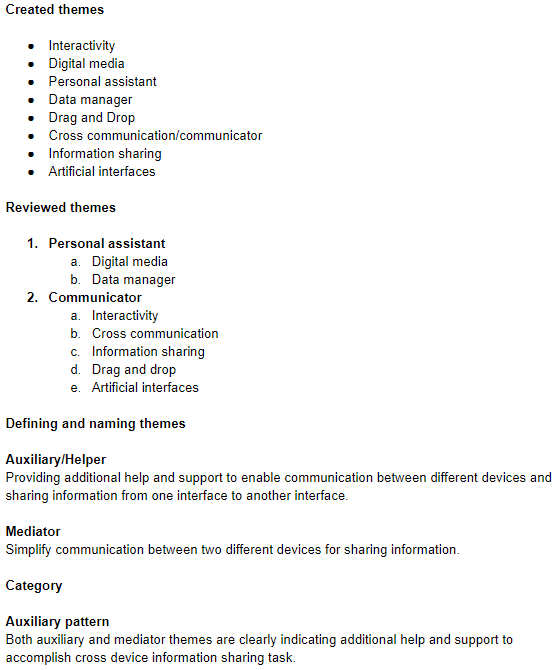}
  \centering
  \caption{Sample output structure of thematic analysis session}
  \label{fig:Thematic_Analysis_Output_Structure}
\end{figure}

\chapter{Themes collection}
\begin{table}[H]
\begin{tabular}{ |p{5.3cm}|p{5.3cm}|p{5.3cm}|  }
\hline
  \textbf{Design patterns} & \textbf{Themes} & \textbf{Categories} \\
  \hline
  MobiES & - Auxiliary/Helper \par - Mediator & Auxiliary pattern \\
  \hline
  The Conduit & - Acquaintance \par - Personalization & Acquaintance/Ally pattern\\
    \hline
  Hyperdrag & - Drag and Drop \par - Controller & Drag and Drop \\
    \hline
  Cross-Device Drag-and-Drop & - Drag and Drop \par - Gesture Control \par - AI Messenger & Gesture Control \\
    \hline
  VisPorter & - Collaborative interaction \par - Gesture Control \par - Proxy & Collaborative interaction \\
    \hline
  MultiSpace & - Portals \par - Collaborative interaction \par - Interactiveness & Portals Interactiveness \\
    \hline
  Conductor & -  Engaging media \par - Gesture Control & Interactor \\
    \hline
  Stitching & - Connector \par - Context aware interaction \par - Gesture Control & Connector \\
    \hline
\end{tabular}
\caption{Document structure of themes collection}
\label{Themes_Collection}
\end{table}

\cleardoublepage
\bibliographystyle{apacite}
\bibliography{references}

\end{document}

%% file: template/thesis.preamble.tex
\usepackage{graphicx}
\usepackage{verbatim}
\usepackage{latexsym}
\usepackage{template/mathchars}
\usepackage{setspace}

\input{template/blocked.sty}
\input{template/uhead.sty}
\input{template/boxit.sty}
\input{template/icthesis.sty}

\newcommand{\NN}{{\sf I\kern-0.14emN}}   
\newcommand{\ZZ}{{\sf Z\kern-0.45emZ}}   
\newcommand{\QQQ}{{\sf C\kern-0.48emQ}}   
\newcommand{\RR}{{\sf I\kern-0.14emR}}   

\newcommand{\normallinespacing}{\renewcommand{\baselinestretch}{1.5} \normalsize}

\newcommand{\syncc}{~\stackrel{\textstyle \rhd\kern-0.57em\lhd}{\scriptstyle L}~}